\documentclass[journal=jacsat,manuscript=article,layout=onecolumn]{achemso}
\usepackage{graphicx}
\usepackage{multicol}

\usepackage[version=3]{mhchem} 
\usepackage{xcolor}
\usepackage{parskip}
\usepackage{hyperref}

\definecolor{HarvardRed}{cmyk}{0.10,1,0.84,0.47}

\usepackage{lineno}

\usepackage{pdfpages}
\author{Jelena Wohlwend}
\email{jelena.wohlwend@mat.ethz.ch}
 \affiliation{Laboratory for Nanometallurgy, Department of Materials, ETH Zurich, 8093 Zürich, Switzerland}
 \author{Anna Hilti}%
 \affiliation{Laboratory for Nanometallurgy, Department of Materials, ETH Zurich, 8093 Zürich, Switzerland}
 \author{Claudiadele Polinari}%
 \affiliation{Laboratory for Nanometallurgy, Department of Materials, ETH Zurich, 8093 Zürich, Switzerland}
\author{Ralph Spolenak}
\affiliation{Laboratory for Nanometallurgy, Department of Materials, ETH Zurich, 8093 Zürich, Switzerland}
 \author{Henning Galinski}%
 \affiliation{Laboratory for Nanometallurgy, Department of Materials, ETH Zurich, 8093 Zürich, Switzerland}
\title[An \textsf{achemso} demo]
  {\textbf{Hybrid resonant metasurfaces with configurable structural colors}}

\abbreviations{IR,visible}

\begin{document}

\begin{abstract}
Metasurfaces play a key role in functionalizing light at the nanoscale. Existing dielectric metasurfaces, however, are often limited to geometric primitives and their usage in emergent hybrid metasurfaces is hampered as confinement of light occurs only in their interior. Taking inspiration from biophotonic systems in nature, we introduce a new class of hybrid metasurfaces, which combine ordered and disordered elements. While the ordered phase relies on non-reciprocal meta-atoms - whose breaking of the out-of-plane symmetry enables the confinement of visible light in air, the disordered phase exploits global plasmonic network modes and their ability to localize energy at nanometric scales. By generating configurable structural colors with extra-ordinary resolution, we demonstrate that coupling of these elements provides a new dimension in the design space. We showcase that control of the local light-matter interaction enables the creation of intricate, customizable optical patterns, which open new avenues for information encoding and high-security features.

\end{abstract}

\section{Introduction}
\textit{"Color is crucial"} - Pop artist Roy Lichtenstein's famous quote highlights the importance of color in art and culture, but the significance of color goes far beyond. In nature color is a means of communication, natural selection~\cite{ontheroleofcolourintheevolutionofsexualselection}, camouflage~\cite{theadaptivevalueofcamouflageandcolourchange} and self-defense~\cite{predation}. From the synthesis of Prussian blue -- the first synthetic pigment -- to the creation of quantum dots -- used in today's display technology, colors and their creation have always reflected scientific advances and the current state of technology. Even though colors bear many complex functions, at their core they emerge simply from the interaction of light with matter.
\par 
This is most tangible in the case of structural colors, where coloration arises from the interaction of light with architectures of a size similar to the incident wavelength. Technologically, structural colors offer various advantages, such as sustainable manufacturing processes, long-term durability, and great versatility that allows for a broad range of applications starting from anti-counterfeiting measures~\cite{anticounterfeitmeasures}, adaptable camouflage~\cite{tunablestructuralcolors}, to color filters~\cite{colorfilters_broadbandabsorbers} and sensors~\cite{bioinspiredstructuralcolosensor,largescale_widegamutcoloration_atdiffractionlimit}. A multitude of structural colors are found in nature, evident in the blue fruits of \textit{Viburnum tinus}~\cite{blueberries}, black feathers of birds of paradise~\cite{birdsofparadies}, the white skin of pyjama squids~\cite{pijamasquid}, and the white scale of the Cyphochilus beetle~\cite{beetles}. In these prominent examples, the unique looks of these fruits and animals arise from a broadband optical response of subwavelength structures in the skin, feathers, and scales.
\par 
Intriguingly, many systems optimized by natural evolution rely on disorder as a design principle to attain specific functions~\cite{disorderinnature}. 
An example thereof is the male peacock spider, where hierarchical disorder of ultra-dense gratings is responsible for their angle-independent coloration used to attract suitable mates~\cite{peacockspider}. Another example is the blue berries of \textit{Viburnum tinus} where the striking metallic blue coloration stems from disordered multilayers of a biphasic structure of lipids and cellulose cell walls~\cite{blueberries}.
\par 
While nature readily utilizes complex photonic structures comprising ordered and disordered elements, nanophotonic systems are often limited to ordered systems. Although initial efforts have been made to embrace disorder as a design principle~\cite{disorderinducedmaterialstefanmaier,manipulatingdisorder,scalableultraresistantstructuralcolorHenning, structuralcolarmonoharan}, so far coupled systems that contain ordered and disordered elements, which enable local control over the light-matter interaction, remain to be explored. This unexplored aspect in photonics, stems from the general complexity of disordered system and the limited availability of ordered systems containing meta-atoms which break the out-of-plane symmetry. 
\begin{figure*}[t!]
\includegraphics[width=1\textwidth]{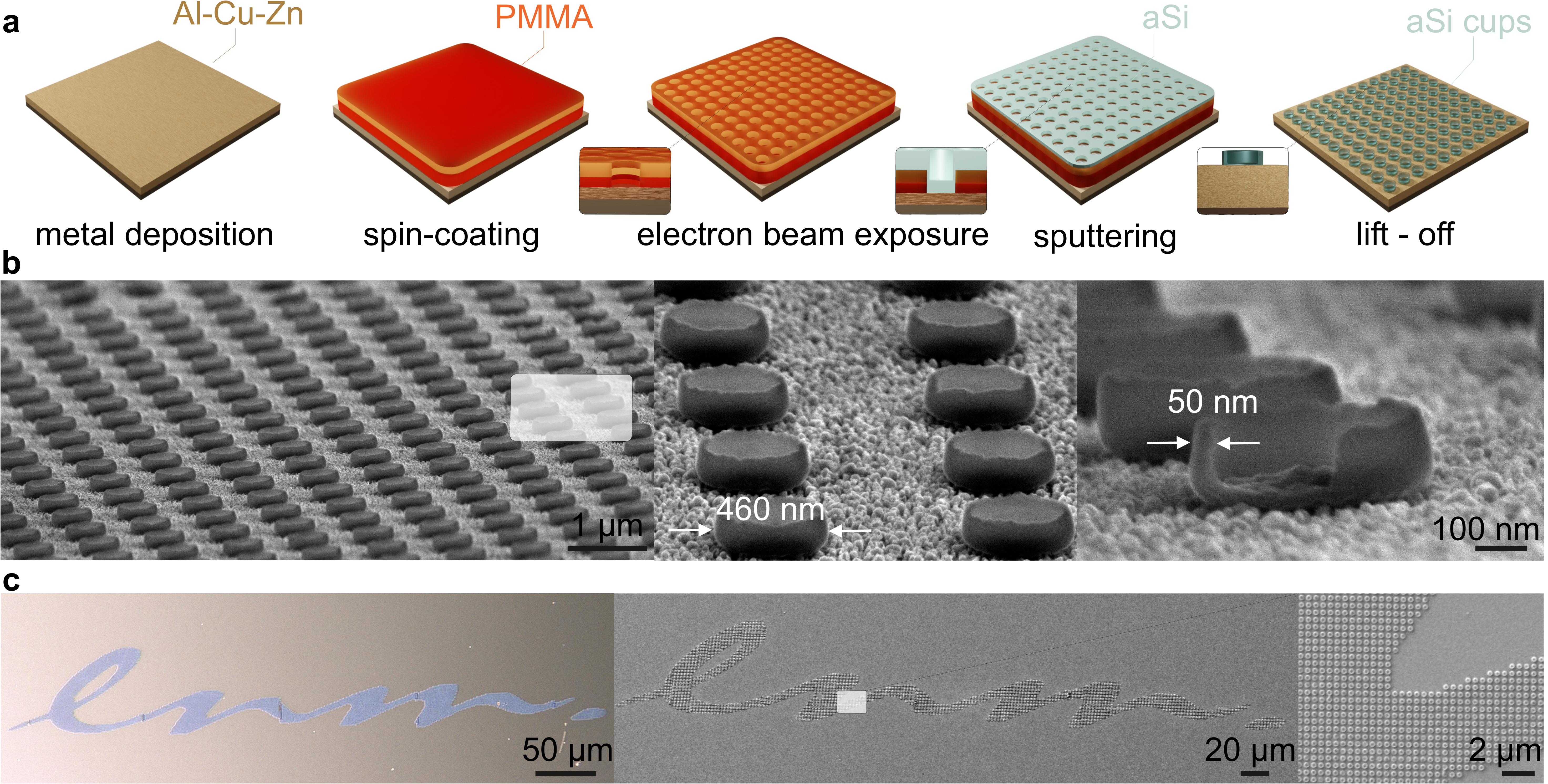}
\caption{
\textbf{Directed growth of amorphous silicon nanocups} 
\textbf{(a)} Scheme depicting the fabrication process of nanocup metasurfaces including (i) metal deposition (ii) spin coating of double layer resist (iii) electron beam exposure (iv) directed growth of a-Si via sputtering (v) lift-off. \textbf{(b)} Selected scanning electron micrographs of the a-Si nanocup arrays highlighting: open resonator structure, high uniformity, and nanometric character (side walls = $50$~nm). \textbf{(c)} Optical micrograph of a optical pattern illustrating the formation of vibrant colors alongside scanning electron micrographs showing the Moiré effect of the nanocup array.}
\label{fig:color_one} 
\end{figure*}
\par 
In this work, we present amorphous silicon (a-Si) nanocups as a versatile and scalable platform to design hybrid metasurfaces (ordered dielectric element + disordered metallic element). In these hybrid system, the out-of-plane asymetry of the dielectric nanocups enables efficient coupling to the disordered metallic photonic element.
\par 
Silicon nanostructures are key building blocks of modern photonics and their ability to strongly enhance and shape light fields at the nanoscale leads to a plethora of applications from non-linear on-chip photonics~\cite{onchipphotonics}, microcavities~\cite{metasurfacecapasso2023} and sensing~\cite{siliconnanostructuressensing} to structural colors~\cite{alldielectricmetasrfaceforhighperformancestructuralcolor}. 
Here, at variance with the widely used geometric primitive dielectric metasurfaces (such as disc arrays), the nanocups constitute open-resonators where the localized optical modes reside in air. Such preferential confinement of light in a lossless medium (air) enables efficient exchange of energy with other resonant systems, such as plasmonic network metamaterials or quantum dots, by radiative coupling~\cite{mievoids}. 
\par 
We fabricate non-primitive a-Si nanocups by combining geometrical shadowing effects during growth and far-from-equilibrium kinetics. As a proof of concept, we demonstrate the localization of light in these nanocups by generating vibrant structural colors with high spatial resolution. By introducing a disordered optical component in the form of a plasmonic network metamaterial~\cite{hgalinski_lighmanipulation,wohlwend_chemicalengineering,wohlwend2023}, we create a configurable hybrid metasurface whose optical properties rely on the local coupling of Mie-resonances (in a-Si nanocups metasurface) with a set of plasmonic eigenmodes (in network metamaterial).
\begin{figure*}[t!]
\includegraphics[width=1\textwidth]{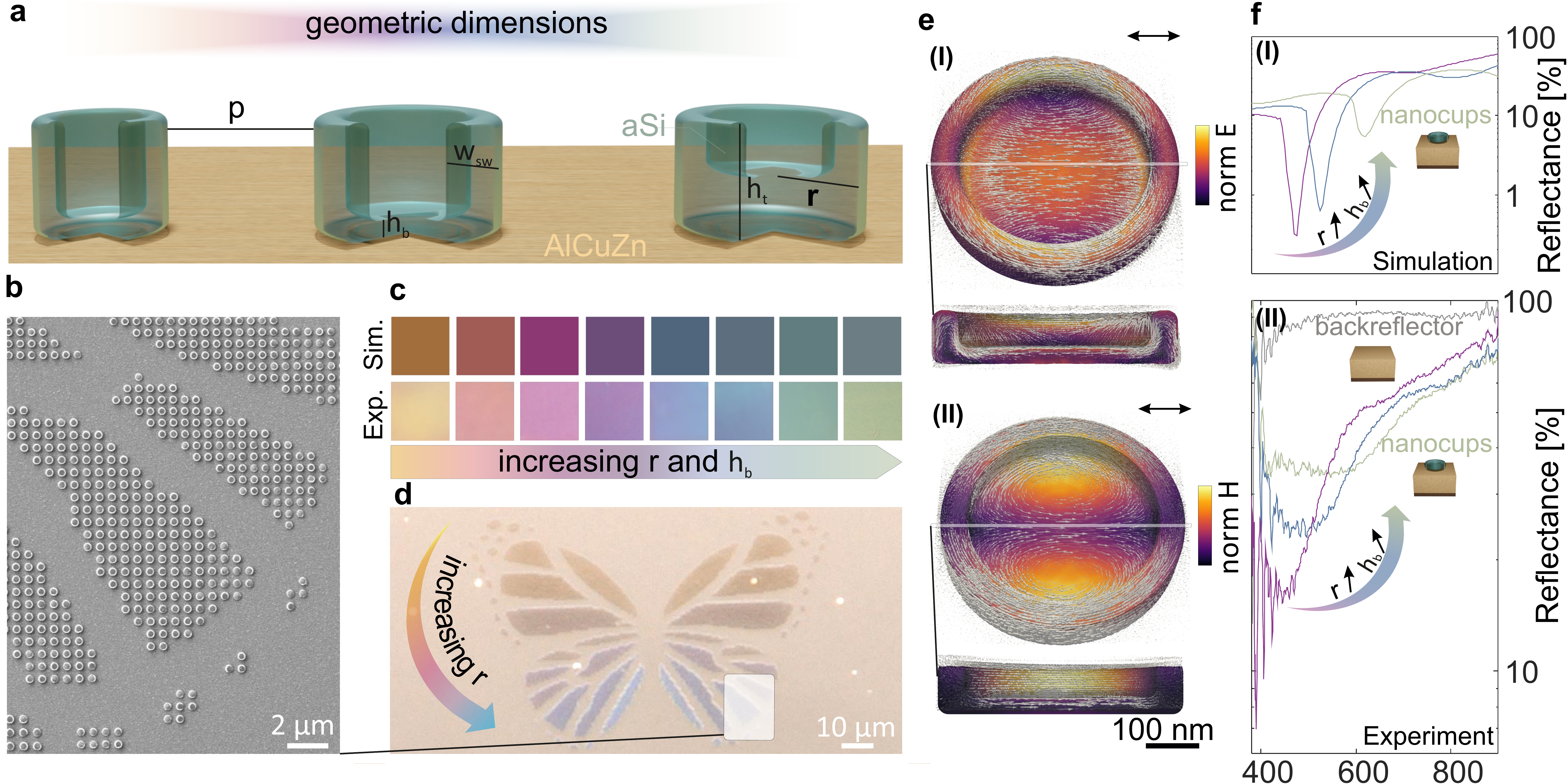}
\caption{\textbf{Local color engineering with a-Si nanocups} \textbf{(a)} Schematic illustration of the color engineering through the change of geometrical parameters. \textbf{(b)} Scanning electron micrograph of a patterned nanocup metasurface. \textbf{(c)} Comparison of simulated colors (from 2D simulations) and experimental colors obtained from optical micrographs of a-Si nanocup metasurfaces, illustrating the continuous color change as function of the radius $r$ and base height $h_b$. \textbf{(d)} Optical micrograph of a-Si nanocups forming an artistic butterfly. The radius of the nanocups increases from top to bottom. \textbf{(e)} FEM simulation of the a-Si nanocup resonator exhibiting an electric dipol resonance (I) and a magnetic dipol resonance (II) with the top view and the cross-section of the normalized electric (I) and magnetic field (II) for the resonance frequencies ($\lambda$ = 370 nm and $\lambda$ = 820 nm). The displacement current is indicated by beige arrows. The black arrow represents the polarization direction of the incident electric field. \textbf{(f)} Simulated and measured reflectance spectra of selected a-Si nanocup metasurfaces showing the formation of an absorbing resonant state.}
\label{fig:color_two} 
\end{figure*}
\section{Results and Discussion}
Nanocup metasurfaces were fabricated using a five-step process (Figure~\ref{fig:color_one}~a) encompassing backreflector deposition, spin-coating, electron beam lithography, nanocup deposition and lift-off. Further details on fabrication are given in the methods section. Both, the aluminum-brass back-reflector and the amorphous silicon (a-Si) nanocups are deposited by physical vapour deposition (PVD). 
\par
The choice of PVD for the fabrication of nanophotonic elements with non-primitive shapes appears counterintuitive at first; as the technique is fundamentally limited to line-of-sight deposition. Thus, when sputtered atoms impinge on non-planar substrates, such as lithographic patterns with nanosized trenches or holes, shadowing effects typically prohibit conformal growth and result in unstable growth dynamics.  
\par 
Still, PVD provides the unique characteristic to deposit under grazing angles. In such a scenario stable highly non-uniform growth can be achieved, provided 
that re-emission effects, due to low wall sticking and high kinetics, are strong enough to redistribute a significant amount of particle flux to otherwise shadowed surfaces~\cite{enhancedstepcoverage,conformalald}.
\par
\begin{figure*}[t!]
\centering
\includegraphics[width=1\textwidth]{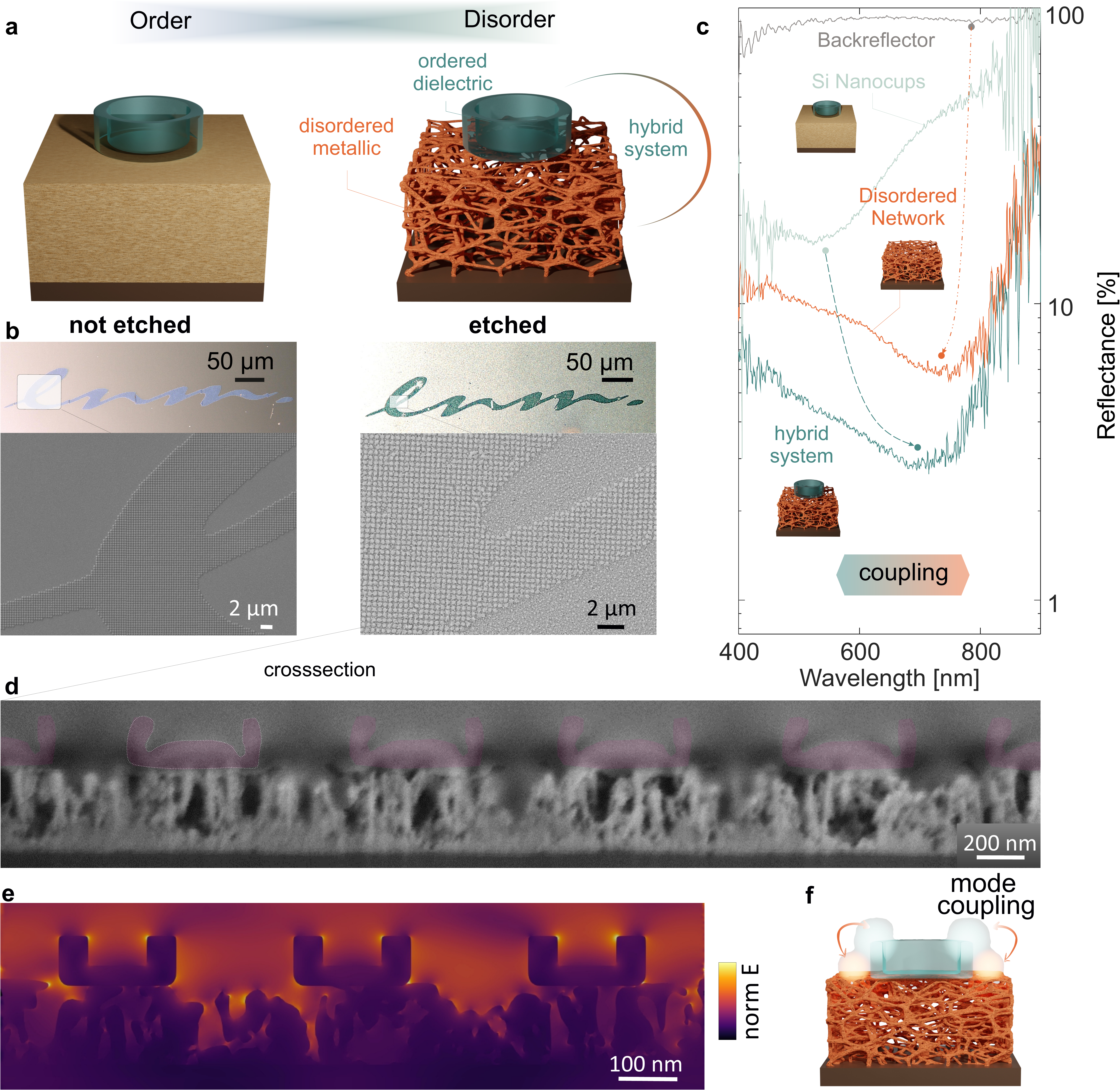}
\caption{ \textbf{Configurable order-disorder transition} 
\textbf{(a)} Schematic illustration of the transition from an ordered to a combined ordered-disordered state. \textbf{(b)} Optical micrographs illustrating the color change through chemical dealloying, i.e., the introduction of disorder in the form of a disordered network metamaterial. \textbf{(c)} Local reflectance measurements of the ordered and disordered states with shifting resonances indicating coupling between the disordered network metamaterial and the nanocup array. \textbf{(d)} FIB-SEM cross-section of the dealloyed state showing the formation of a disordered network out of the aluminum brass. The a-Si nanocups on top of the network (shaded in pink) remain unaffected by the chemical process. \textbf{(e)} Simulated local field enhancement on resonance ($\lambda$= 340 nm) and \textbf{(f)} schematic illustration of the coupling between an a-Si nanocup and "hot spot" of the disordered network metamaterial}
\label{fig:color_three} 
\end{figure*}
A set of nanocup metasurfaces utilizing such directed growth conditions are presented in Figure \ref{fig:color_one} b and c. Here, the systematic optimization of pressure and deposition rate facilitated the tailoring of the a-Si deposition, ensuring complete trench coverage of the inverse nanopattern. Interestingly, predetermined breaking points form at the top of the trenches. We attribute the formation of such breaking points to non-uniform growth conditions resulting in a thickness gradient, with the thinnest wall thickness occurring at the trenches' top (see Supplementary Information).
\par
In Figure \ref{fig:color_one}~b and c, selected sets of nanocups with a diameter of 460~nm  and 320~nm and periodicities of 520~nm (Figure \ref{fig:color_one}~b) and 100~nm  (Figure \ref{fig:color_one}~c) are shown. The nanocup walls are 47~nm thick and show a slight tapering towards the bottom ($\alpha = 8^{\circ}$), which results from the point spread function of the electron beam lithography process. Colored lettering can be observed as the a-Si nanocups are arranged in a predefined pattern (Figure \ref{fig:color_one}~c) confirming the localization of light in the metasurface. Strikingly, the high uniformity of the a-Si nanocup metasurface leads to the formation of Moiré patterns in the scanning electron micrograph (Figure~\ref{fig:color_one} c), affirming the robustness of the chosen fabrication route.
\par 
The nanocups provide great design freedom enabling an elaborate control of the resonant optical modes ~\cite{controlingmagneticandelectricdipolemodes,engineeringlightabsorpitioninsemicon}. Compared to  nanorods or nanodisks, nanocups have additional geometric parameters such as base height h$_b$, sidewall width s$_{sw}$, height $h_{\text{t}}$ which can be tailored to control the resulting light–matter interaction (Figure \ref{fig:color_two}~a). Additionally, the out-of-plane symmetry breaking results in non-reciprocity, i. e., the emergence of different modes along opposite excitation directions (See Supplementary Information)~\cite{outofplanesymmetrybreaking}. To illustrate the full potential of the a-Si nanocups, we perform a set of full-wave simulations on 2D projections of the nanocups with changing geometric dimensions to study their impact on the resonant behavior.
\par 
\begin{figure*}[t!]
\centering
\includegraphics[width=1\textwidth]{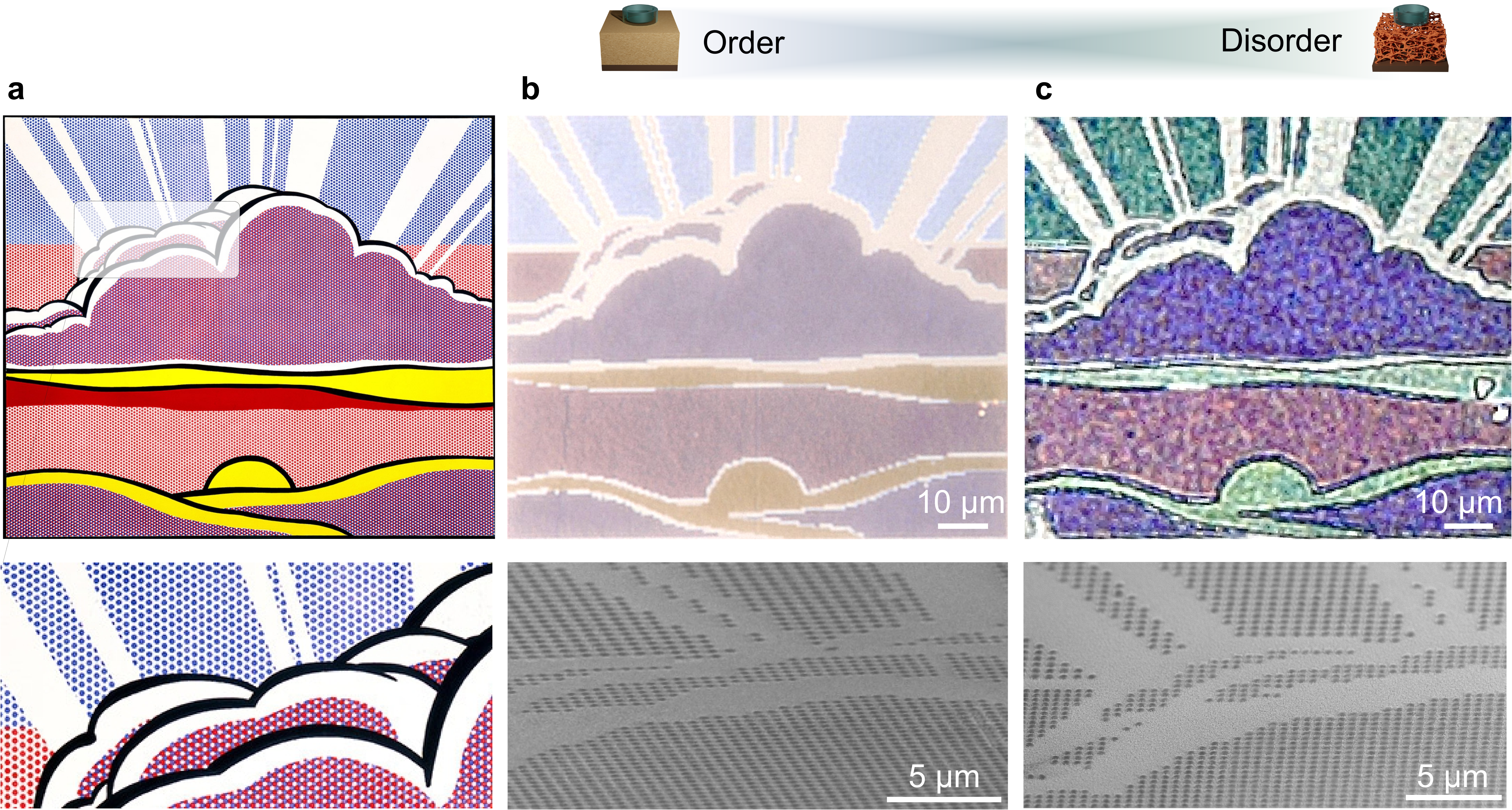}
\caption{
 \textbf{Generating Colors with Points} \textbf{(a)} Roy Lichtensteins "Setting Sun" \textbf{(b)} optical micrograph of nanoscale reproduction of the "Setting sun" with a-Si nanocup metasurface \textbf{(c)}  Optical micrograph illustrating the color change through chemical dealloying, i.e., the formation of a disordered optical cavity and consequent coupling between the aSi nanocups and the disordered cavity. The subfigures at the bottom present the color centers, i. e. the points generating the image.}
 \label{fig:color_four}
\end{figure*}
The simulated optical response of the a-Si nanocups is marked by a Lorentz-like dip in reflection (Figure \ref{fig:color_two}~f~I). This resonance exhibits quasi-perfect attenuation of light and counts responsible for the perceived coloration. By modelling the optical scattering of an isolated nanocup, we can relate such resonant behavior to the emergence of an electric dipolar (ED) mode in the nanocup (Figure \ref{fig:color_two}~e). While the normalized electric field of the mode aligns with the incident field, light is confined between the cup side walls in air (Figure \ref{fig:color_two}~e). Such confinement in air is beneficial as the mode is not impacted by damping and is in principle able to couple to other resonant systems in the near field. At higher wavelength a magnetic dipole mode (MD) appears due to  circular displacement currents in the base of the nanocup (Figure \ref{fig:color_two}~f). 
When arranged in an array and placed on a backreflector, the metasurface also exhibits a surface lattice resonance (SLR) which is confined between resonators and strongly depends on the chosen periodicity $p$ (See Supplementary Information).
\par
Moreover, 3D simulations of the scattering behavior of the a-Si nanocups reveal predominantly forward scattering (see Supplementary Information) enhancing the interaction between nanocups and the substrate. This is especially advantageous for efficient coupling to a second resonating system.
\par 
With increasing cup diameter $2r$ the absorbance resonance undergoes a red shift resulting in the formation of a full gamut of structural colors (Figure \ref{fig:color_two}~c and Supplementary Information). The calculated colors based on the 2D simulations are shown in Figure \ref{fig:color_two}~c. Additionally to the cup diameter also the periodicity $p$, the base height h$_b$ and the side wall thickness s$_{sw}$ were investigated and have been found to influence the resulting colors (see Supplementary Information).
\par
We continue with the fabrication of a-Si nanocup metasurfaces with increasing cup diameter, shown in Figure \ref{fig:color_two}~c and d. In agreement with simulations, the experimentally observed resonance of the nanocups red-shifts for an increase in cup diameter (Figure \ref{fig:color_two}~f~II) and is accompanied by a broadening of the resonances. The resulting structural colors with increasing diameter are shown in Figure \ref{fig:color_two}~c and match the simulations under the condition that an increase in cup diameter $r$ is accompanied by an increase in base height $h_{b}$ (see Supplementary Information). The observed difference in the resonant line width between the simulation and experiments (Figure \ref{fig:color_two}~e) most likely stems from the difference in light collection and additionally from imperfections of the a-Si nanocups. While in the simulation the normal incident reflection is calculated, in the experiments an objective is employed to locally measure the reflected light. The numerical aperture of the objective results in a spread of collection angles responsible for the observed broadening of the optical response.
\par
Drawing inspiration from the multitude of structural colors in butterflies \cite{structualcolors_butterflies,replicatingstructuralcolorsinbutterlies}, we engineer a butterfly of our own (Figure \ref{fig:color_two}~d). The different segments of the butterfly wings contain a-Si nanocups of different diameters, base heights and periodicities. The structural colors highlight the versatility, scalability and local color control down to a single pixel, i.e. a single nanocup (see Figure~\ref{fig:color_two}~b).
\par 
Having established a resonant non-reciprocal meta-atom by breaking the out-of-plane symmetry, we can now continue and introduce a disordered photonic element. We thereby reconfigure the backreflector through chemical dealloying and create a hybrid metasurface consisting of a-Si nanocups and a disordered network metamaterial (DNM) (Figure~\ref{fig:color_three}~a). DNMs are able to trap and localize surface plasmon (SP) waves in nanometric volumes, so-called "hot spots". Thus, in contrast to simple nanoparticles, the optical response of such a network takes on a global character and can be conceptualized as a network of randomly coupled dipoles~\cite{Chen2018,hgalinski_lighmanipulation}. Chemical engineering of the initial system allows to control the network architecture and topology - critical parameters to design the resonant modes in the system~\cite{wohlwend_chemicalengineering}. Due to their high hot spot density and enhanced modal overlap, DNMs exhibit strong coupling to other photonic elements, such as nanoparticles~\cite{wohlwend2023}, rendering them ideal for hybrid optical systems. 
\par 
Using basic wet-chemistry, the plasmonic disordered network can be assembled \textit{in-place}, i.e. using the existing metasurface, by chemical dealloying of the aluminum brass backreflector. Chemical dealloying — a self-organizing process — creates an open porous nanometric network with a variety of different curvatures and local variations in edge length and connectivity (Figure~\ref{fig:color_three}~d)~\cite{wohlwend_chemicalengineering,wohlwend2023}. While the a-Si nanocups remain unaffected by the chemical process, the backreflector undergoes a drastic change from a reflective metallic substrate to disordered metamaterial with resonant modes (Figure~\ref{fig:color_three}~b and d). 
\par
To analyze the impact of the order-disorder transition on the local optical properties, we compare the local reflectance of the ordered system (a-Si nanocups + backreflector) with the disordered system (a-Si nanocups + DNM). Qualitatively, the change is visible in Figure \ref{fig:color_three}~b as a color shift of the optical pattern from blue to green. This color change matches the reflectance spectra (Figure~\ref{fig:color_three}~c), where a red shift of the resonance is observed (dashed blue arrow). In the presence of the nanocups, the absorption is enhanced. Additionally, comparing the backreflector and the disordered network, the formation of a resonant plasmonic mode characterized by the emergence of a reflectance minimum is observed (Figure \ref{fig:color_three}~d dot-dot-dashed orange arrow). 
\par 
Both the change in coloration and the shift of the resonant modes are evocative of mode coupling. Full wave simulations of the resonant state (Figure~\ref{fig:color_three}~e) furthermore show the local field enhancement, indicating the  formation of a combined optical state. This combined state results from highly localized energy exchange between ordered and disordered photonic elements. Here the interplay of the out-of-plane asymmetry of the ordered elements (a-Si nanocups) and the disorder (DNM), resulting in the equipartition of energy between all plasmonic modes, enables configurable local control over the light–matter interactions~\cite{wohlwend2023,manipulatingdisorder}.
\par 
To illustrate such configurability between an ordered and a disordered state, we reproduced Roy Lichtensteins "Sinking Sun" (1964), shown in Figure \ref{fig:color_four}~a. Similar to popular impressionist paintings, Lichtensteins paintings - consisting of different colored points with defined spacings between them - are most legible at a distance and increasingly abstract upon enlargement. In our reproductions the colors are made by point-like structures, i.e. the a-Si nanocups arranged in a pattern constructing the image. Remarkably, even single lines of a-Si nanocups generate clearly defined colors. Across the order-disorder transition, the metasurface experiences a drastic change in the local structural coloration. While in the ordered state a pastel color palette dominates, the disordered state exhibits highly saturated structural colors. Such transition underpins the change of the local-light matter interaction, that relies on coupling between a-Si nanocup metasurface and the DNM.
\par 
The aforementioned nanoscale reproduction of the artwork demonstrates the creation of intricate and customizable optical patterns by our hybrid metasurface. Furthermore, the ability to engineer specific resonant states through geometric dimensions and chemical compositions extends the design space and  makes our systems ideal for encoding information in a visually imperceptible manner. This special property enables for example the fabrication of high-security features that are challenging to replicate, providing an effective deterrent against counterfeiting. 

\section{Conclusion}
In conclusion, we have developed new class of hybrid metasurfaces based on ordered and disordered photonic elements. The ordered dielectric nanocup metasurfaces are fabricated by directed growth, where we take advantage of the non-uniform growth conditions of PVD to create the out-of-plane asymmetrical nanostructures. This simple fabrication scheme is not restricted to a-Si and thus can readily be employed to other semiconductors such as germanium or titanium dioxid. We showcase efficient and robust light absorption in the non-reciprocal nanocup metasurfaces through the generation of a full gamut of structural colors. Beyond the generation of structural colors these open resonators offer unique possibilities for applied photonics including lasing and colloidal quantum dot electronics as the cups in principle, can be filled with a gain medium. 
Furthermore, the hybrid metasurface introduced here can serve as a testbed to study the light–matter interaction and mode coupling of ordered and disordered components in complex photonic systems. Especially, interesting would be study the mode hybridisation of such systems in three dimensions by energy electron loss spectroscopy (EELS) tomography.

\section{Experimental Section}
\textbf{\textit{Thin-Film Deposition:}} The aluminum brass backreflector, was deposited by direct DC magnetron sputtering onto $Si/SiO_2$ substrates. A detailed description on the selection of the ideal aluminum brass for this purpose is given in the supplementary material. 
\par 
\textbf{\textit{Nanocup fabrication:}} The a-Si nanocups were fabricated by electron-beam exposure, carried out using a RaithOne system operating at 30 keV acceleration voltage. The inverse nano-pattern was generated within a double-layer PMMA electron beam resist and developed in IPA/MIBK 3:1. Following the development, Si was deposited by magnetron sputtering at a rate of 0.3 \r{A}/s with a constant rotation speed of 10 rpm and under a gracing angle of 23.5 degrees.  Finally, a lift off step was performed by immersing the sample in acetone.
\par 
\textbf{\textit{Simulations}}

Three sets of simulations were performed. Firstly, simulations of the periodic arrays of a-Si nanocups were performed using the frequency-domain solver of the commercial finite-element package COMSOL Multiphysics 6.1 in 2D. The a-Si nanocups were modelled by assembling rectangles into a cup shape, were the edges were rounded using the fillet function. Periodic boundary conditions were used for the x-direction to introduce lateral periodicity, while in the upper and lower air boundary of the simulation domain perfectly matched layers (PML) were introduced to absorb the outgoing waves.  The refractive indices of the aluminum brass backreflector and the a-Si were extracted from ellipsometry measurements.

To simulate the scattering spectra of single nanocups again the commercial finite-element package COMSOL Multiphysic 6.1 was used. The cup shape was generated as the difference between two different cylinders. Perfectly matched layers were introduced in spherical shells surrounding the nanocups. The excitation was implemented by a background plane wave. 

Lastly, the introduction of disorder was modeled by introducing a FIB cross-section of the disordered network below the a-Si nanocups in the 2D simulations. The image to curve plugin in Comsol Multiphysics was used to convert the FIB cross-section into a curve.

\textbf{\textit{Color palettes:}} The color palettes as well as the chromaticities were obtained from simulated reflectance spectra using the commercial Software Wolfram Mathematica 12.2. 
\par 
\textbf{\textit{Chemical dealloying:}} Disordered Cu-Zn metamaterial networks were fabricated by chemical dealloying of aluminum brass backreflector. The Al is thereby selectively removed by immersing the films into a 1 M NaOH aqueous solution, where a subwavelength open-porous disordered network is formed~\cite{wohlwend_chemicalengineering}.
\par 
\textbf{\textit{Sample Characterisation:}}
The refractive index as well as angle-dependent reflectance measurements of the thin films were obtained with a M-2000 Woollam ellipsometer. Reflectance spectra of the nanocup metasurfaces were recorded using a custom-built microscopy setup based on an inverse Zeiss Axio Observer.A1m microscope with an Ocean Optics QE-PRO spectrometer and an Ocean Optics Halogen light source HL-2000-FHSA . An aluminum mirror was used as calibration standard.
\par 
\section*{Acknowledgements}
The authors acknowledge the infrastructure and support of FIRST. Electron microscopy analysis was performed at ScopeM, the microscopy platform of ETH Z\"urich. The authors are grateful to Maxence Menétrey for his experimental assistance and his constructive feedback on the manuscript. The authors also thank Marcello Pozzi and José Luis Ocaña Pujol for their help.
\par
\textbf{Author contributions:}
J.W., R.S. and H.G. conceived the research plan. J.W., A.H. and C.P. fabricated the samples. J.W. performed simulations, sample characterisations, data analysis and visualization. J.W. and H.G. wrote the manuscript. All authors reviewed and commented the manuscript.
\par 
\textbf{Competing interests:}
The authors declare that they have no competing interests.

\bibliography{paper.bib}

\providecommand{\latin}[1]{#1}
\makeatletter
\providecommand{\doi}
  {\begingroup\let\do\@makeother\dospecials
  \catcode`\{=1 \catcode`\}=2 \doi@aux}
\providecommand{\doi@aux}[1]{\endgroup\texttt{#1}}
\makeatother
\providecommand*\mcitethebibliography{\thebibliography}
\csname @ifundefined\endcsname{endmcitethebibliography}  {\let\endmcitethebibliography\endthebibliography}{}
\begin{mcitethebibliography}{35}
\providecommand*\natexlab[1]{#1}
\providecommand*\mciteSetBstSublistMode[1]{}
\providecommand*\mciteSetBstMaxWidthForm[2]{}
\providecommand*\mciteBstWouldAddEndPuncttrue
  {\def\EndOfBibitem{\unskip.}}
\providecommand*\mciteBstWouldAddEndPunctfalse
  {\let\EndOfBibitem\relax}
\providecommand*\mciteSetBstMidEndSepPunct[3]{}
\providecommand*\mciteSetBstSublistLabelBeginEnd[3]{}
\providecommand*\EndOfBibitem{}
\mciteSetBstSublistMode{f}
\mciteSetBstMaxWidthForm{subitem}{(\alph{mcitesubitemcount})}
\mciteSetBstSublistLabelBeginEnd
  {\mcitemaxwidthsubitemform\space}
  {\relax}
  {\relax}

\bibitem[Dresp(2009)]{ontheroleofcolourintheevolutionofsexualselection}
Dresp,~B. \emph{On the Role of Colour in the Evolution of Sexual Selection Behaviour}; 2009; pp 110--115\relax
\mciteBstWouldAddEndPuncttrue
\mciteSetBstMidEndSepPunct{\mcitedefaultmidpunct}
{\mcitedefaultendpunct}{\mcitedefaultseppunct}\relax
\EndOfBibitem
\bibitem[Duarte \latin{et~al.}(2018)Duarte, Stevens, and Flores]{theadaptivevalueofcamouflageandcolourchange}
Duarte,~R.; Stevens,~M.; Flores,~A. The adaptive value of camouflage and colour change in a polymorphic prawn. \emph{Scientific Reports} \textbf{2018}, \emph{8}\relax
\mciteBstWouldAddEndPuncttrue
\mciteSetBstMidEndSepPunct{\mcitedefaultmidpunct}
{\mcitedefaultendpunct}{\mcitedefaultseppunct}\relax
\EndOfBibitem
\bibitem[Jeschke \latin{et~al.}(2021)Jeschke, Laforsch, Diel, Diller, Horstmann, and Tollrian]{predation}
Jeschke,~J.; Laforsch,~C.; Diel,~P.; Diller,~J.; Horstmann,~M.; Tollrian,~R. \emph{Predation}; 2021\relax
\mciteBstWouldAddEndPuncttrue
\mciteSetBstMidEndSepPunct{\mcitedefaultmidpunct}
{\mcitedefaultendpunct}{\mcitedefaultseppunct}\relax
\EndOfBibitem
\bibitem[Teutoburg-Weiss \latin{et~al.}(2022)Teutoburg-Weiss, Soldera, Bouchard, Kreß, Vaynzof, and Lasagni]{anticounterfeitmeasures}
Teutoburg-Weiss,~S.; Soldera,~M.; Bouchard,~F.; Kreß,~J.; Vaynzof,~Y.; Lasagni,~A.~F. Structural colors with embedded anti-counterfeit features fabricated by laser-based methods. \emph{Optics \& Laser Technology} \textbf{2022}, \emph{151}, 108012\relax
\mciteBstWouldAddEndPuncttrue
\mciteSetBstMidEndSepPunct{\mcitedefaultmidpunct}
{\mcitedefaultendpunct}{\mcitedefaultseppunct}\relax
\EndOfBibitem
\bibitem[Zhang \latin{et~al.}(2023)Zhang, Tian, Liu, Liu, Zhao, Li, Wang, Chen, and Shao]{tunablestructuralcolors}
Zhang,~W.; Tian,~H.; Liu,~T.; Liu,~H.; Zhao,~F.; Li,~X.; Wang,~C.; Chen,~X.; Shao,~J. Chameleon-inspired active tunable structural color based on smart skin with multi-functions of structural color{,} sensing and actuation. \emph{Mater. Horiz.} \textbf{2023}, \emph{10}, 2024--2034\relax
\mciteBstWouldAddEndPuncttrue
\mciteSetBstMidEndSepPunct{\mcitedefaultmidpunct}
{\mcitedefaultendpunct}{\mcitedefaultseppunct}\relax
\EndOfBibitem
\bibitem[Ji \latin{et~al.}(2017)Ji, Lee, Xu, Zhou, Park, and Guo]{colorfilters_broadbandabsorbers}
Ji,~C.; Lee,~K.-T.; Xu,~T.; Zhou,~J.; Park,~H.~J.; Guo,~L.~J. Engineering Light at the Nanoscale: Structural Color Filters and Broadband Perfect Absorbers. \emph{Advanced Optical Materials} \textbf{2017}, \emph{5}, 1700368\relax
\mciteBstWouldAddEndPuncttrue
\mciteSetBstMidEndSepPunct{\mcitedefaultmidpunct}
{\mcitedefaultendpunct}{\mcitedefaultseppunct}\relax
\EndOfBibitem
\bibitem[Qin \latin{et~al.}(2019)Qin, Sun, Hua, and He]{bioinspiredstructuralcolosensor}
Qin,~M.; Sun,~M.; Hua,~M.; He,~X. Bioinspired structural color sensors based on responsive soft materials. \emph{Current Opinion in Solid State and Materials Science} \textbf{2019}, \emph{23}, 13--27, Active and adaptive soft matter\relax
\mciteBstWouldAddEndPuncttrue
\mciteSetBstMidEndSepPunct{\mcitedefaultmidpunct}
{\mcitedefaultendpunct}{\mcitedefaultseppunct}\relax
\EndOfBibitem
\bibitem[Li \latin{et~al.}(2022)Li, Xiang, Elizarov, Makarenko, Lopez, Getman, Bonifazi, Mazzone, and Fratalocchi]{largescale_widegamutcoloration_atdiffractionlimit}
Li,~N.; Xiang,~F.; Elizarov,~M.; Makarenko,~M.; Lopez,~A.; Getman,~F.; Bonifazi,~M.; Mazzone,~V.; Fratalocchi,~A. Large‐Scale and Wide‐Gamut Coloration at the Diffraction Limit in Flexible, Self‐Assembled Hierarchical Nanomaterials. \emph{Advanced Materials} \textbf{2022}, \emph{34}, 2108013\relax
\mciteBstWouldAddEndPuncttrue
\mciteSetBstMidEndSepPunct{\mcitedefaultmidpunct}
{\mcitedefaultendpunct}{\mcitedefaultseppunct}\relax
\EndOfBibitem
\bibitem[Middleton \latin{et~al.}(2020)Middleton, Sinnott-Armstrong, Ogawa, Jacucci, Moyroud, Rudall, Prychid, Conejero, Glover, Donoghue, and Vignolini]{blueberries}
Middleton,~R.; Sinnott-Armstrong,~M.; Ogawa,~Y.; Jacucci,~G.; Moyroud,~E.; Rudall,~P.~J.; Prychid,~C.; Conejero,~M.; Glover,~B.~J.; Donoghue,~M.~J.; Vignolini,~S. Viburnum tinus Fruits Use Lipids to Produce Metallic Blue Structural Color. \emph{Current Biology} \textbf{2020}, \emph{30}, 3804--3810.e2\relax
\mciteBstWouldAddEndPuncttrue
\mciteSetBstMidEndSepPunct{\mcitedefaultmidpunct}
{\mcitedefaultendpunct}{\mcitedefaultseppunct}\relax
\EndOfBibitem
\bibitem[McCoy \latin{et~al.}(2018)McCoy, Feo, Harvey, and Prum]{birdsofparadies}
McCoy,~D.; Feo,~T.; Harvey,~T.; Prum,~R. Structural absorption by barbule microstructures of super black bird of paradise feathers. \emph{Nature Communications} \textbf{2018}, \emph{9}\relax
\mciteBstWouldAddEndPuncttrue
\mciteSetBstMidEndSepPunct{\mcitedefaultmidpunct}
{\mcitedefaultendpunct}{\mcitedefaultseppunct}\relax
\EndOfBibitem
\bibitem[Bell \latin{et~al.}(2014)Bell, Mathger, Gao, Senft, Kuzirian, Kattawar, and Hanlon]{pijamasquid}
Bell,~G.; Mathger,~L.; Gao,~M.; Senft,~S.; Kuzirian,~A.; Kattawar,~G.; Hanlon,~R. Diffuse White Structural Coloration from Multilayer Reflectors in a Squid. \emph{Advanced materials (Deerfield Beach, Fla.)} \textbf{2014}, \emph{26}\relax
\mciteBstWouldAddEndPuncttrue
\mciteSetBstMidEndSepPunct{\mcitedefaultmidpunct}
{\mcitedefaultendpunct}{\mcitedefaultseppunct}\relax
\EndOfBibitem
\bibitem[Vukusic \latin{et~al.}(2007)Vukusic, Hallam, and Noyes]{beetles}
Vukusic,~P.; Hallam,~B.; Noyes,~J. Brilliant Whiteness in Ultrathin Beetle Scales. \emph{Science (New York, N.Y.)} \textbf{2007}, \emph{315}, 348\relax
\mciteBstWouldAddEndPuncttrue
\mciteSetBstMidEndSepPunct{\mcitedefaultmidpunct}
{\mcitedefaultendpunct}{\mcitedefaultseppunct}\relax
\EndOfBibitem
\bibitem[Rothammer \latin{et~al.}(2021)Rothammer, Zollfrank, Busch, and von Freymann]{disorderinnature}
Rothammer,~M.; Zollfrank,~C.; Busch,~K.; von Freymann,~G. Tailored Disorder in Photonics: Learning from Nature. \emph{Advanced Optical Materials} \textbf{2021}, \emph{9}, 2100787\relax
\mciteBstWouldAddEndPuncttrue
\mciteSetBstMidEndSepPunct{\mcitedefaultmidpunct}
{\mcitedefaultendpunct}{\mcitedefaultseppunct}\relax
\EndOfBibitem
\bibitem[Wilts \latin{et~al.}(2020)Wilts, Otto, and Stavenga]{peacockspider}
Wilts,~B.~D.; Otto,~J.; Stavenga,~D.~G. Ultra-dense{,} curved{,} grating optics determines peacock spider coloration. \emph{Nanoscale Adv.} \textbf{2020}, \emph{2}, 1122--1127\relax
\mciteBstWouldAddEndPuncttrue
\mciteSetBstMidEndSepPunct{\mcitedefaultmidpunct}
{\mcitedefaultendpunct}{\mcitedefaultseppunct}\relax
\EndOfBibitem
\bibitem[Mao \latin{et~al.}(2021)Mao, Liu, Niu, Qin, Song, Han, Palmer, Maier, and Zhang]{disorderinducedmaterialstefanmaier}
Mao,~P.; Liu,~C.; Niu,~Y.; Qin,~Y.; Song,~F.; Han,~M.; Palmer,~R.~E.; Maier,~S.~A.; Zhang,~S. Disorder-Induced Material-Insensitive Optical Response in Plasmonic Nanostructures: Vibrant Structural Colors from Noble Metals. \emph{Advanced Materials} \textbf{2021}, \emph{33}, 2007623\relax
\mciteBstWouldAddEndPuncttrue
\mciteSetBstMidEndSepPunct{\mcitedefaultmidpunct}
{\mcitedefaultendpunct}{\mcitedefaultseppunct}\relax
\EndOfBibitem
\bibitem[Mao \latin{et~al.}(2020)Mao, Liu, Song, Han, Maier, and Zhang]{manipulatingdisorder}
Mao,~P.; Liu,~C.; Song,~F.; Han,~M.; Maier,~S.; Zhang,~S. Manipulating disordered plasmonic systems by external cavity with transition from broadband absorption to reconfigurable reflection. \emph{Nature Communications} \textbf{2020}, \emph{11}\relax
\mciteBstWouldAddEndPuncttrue
\mciteSetBstMidEndSepPunct{\mcitedefaultmidpunct}
{\mcitedefaultendpunct}{\mcitedefaultseppunct}\relax
\EndOfBibitem
\bibitem[Galinski \latin{et~al.}(2016)Galinski, Favraud, Dong, Totero~Gongora, Favaro, Döbeli, Spolenak, Fratalocchi, and Capasso]{scalableultraresistantstructuralcolorHenning}
Galinski,~H.; Favraud,~G.; Dong,~H.; Totero~Gongora,~J.~S.; Favaro,~G.; Döbeli,~M.; Spolenak,~R.; Fratalocchi,~A.; Capasso,~F. Scalable, ultra-resistant structural colors based on network metamaterials. \emph{Light: Science \& Applications} \textbf{2016}, \emph{6}, e16233\relax
\mciteBstWouldAddEndPuncttrue
\mciteSetBstMidEndSepPunct{\mcitedefaultmidpunct}
{\mcitedefaultendpunct}{\mcitedefaultseppunct}\relax
\EndOfBibitem
\bibitem[Hwang \latin{et~al.}(2021)Hwang, Stephenson, Barkley, Brandt, Xiao, Aizenberg, and Manoharan]{structuralcolarmonoharan}
Hwang,~V.; Stephenson,~A.; Barkley,~S.; Brandt,~S.; Xiao,~M.; Aizenberg,~J.; Manoharan,~V. Designing angle-independent structural colors using Monte Carlo simulations of multiple scattering. \emph{Proceedings of the National Academy of Sciences} \textbf{2021}, \emph{118}, e2015551118\relax
\mciteBstWouldAddEndPuncttrue
\mciteSetBstMidEndSepPunct{\mcitedefaultmidpunct}
{\mcitedefaultendpunct}{\mcitedefaultseppunct}\relax
\EndOfBibitem
\bibitem[Xie \latin{et~al.}(2022)Xie, Xiang, Chang, Jin, Peters, and Bowers]{onchipphotonics}
Xie,~W.; Xiang,~C.; Chang,~L.; Jin,~W.; Peters,~J.; Bowers,~J.~E. Silicon-integrated nonlinear III-V photonics. \emph{Photon. Res.} \textbf{2022}, \emph{10}, 535--541\relax
\mciteBstWouldAddEndPuncttrue
\mciteSetBstMidEndSepPunct{\mcitedefaultmidpunct}
{\mcitedefaultendpunct}{\mcitedefaultseppunct}\relax
\EndOfBibitem
\bibitem[Ossiander \latin{et~al.}(2023)Ossiander, Meretska, Rourke, Spägele, Yin, Benea-Chelmus, and Capasso]{metasurfacecapasso2023}
Ossiander,~M.; Meretska,~M.; Rourke,~S.; Spägele,~C.; Yin,~X.; Benea-Chelmus,~I.-C.; Capasso,~F. Metasurface-stabilized optical microcavities. \emph{Nature Communications} \textbf{2023}, \emph{14}\relax
\mciteBstWouldAddEndPuncttrue
\mciteSetBstMidEndSepPunct{\mcitedefaultmidpunct}
{\mcitedefaultendpunct}{\mcitedefaultseppunct}\relax
\EndOfBibitem
\bibitem[Vendamani \latin{et~al.}(2022)Vendamani, Rao, Pathak, and Soma]{siliconnanostructuressensing}
Vendamani,~V.~S.; Rao,~S. V. S.~N.; Pathak,~A.~P.; Soma,~V.~R. Silicon Nanostructures for Molecular Sensing: A Review. \emph{ACS Applied Nano Materials} \textbf{2022}, \emph{5}, 4550--4582\relax
\mciteBstWouldAddEndPuncttrue
\mciteSetBstMidEndSepPunct{\mcitedefaultmidpunct}
{\mcitedefaultendpunct}{\mcitedefaultseppunct}\relax
\EndOfBibitem
\bibitem[Wenhong \latin{et~al.}(2020)Wenhong, Xiao, Song, Liu, Wu, Wang, Yu, Han, and Tsai]{alldielectricmetasrfaceforhighperformancestructuralcolor}
Wenhong,~Y.; Xiao,~S.; Song,~Q.; Liu,~Y.; Wu,~Y.; Wang,~S.; Yu,~J.; Han,~J.; Tsai,~D.~P. All-dielectric metasurface for high-performance structural color. \emph{Nature Communications} \textbf{2020}, \emph{11}\relax
\mciteBstWouldAddEndPuncttrue
\mciteSetBstMidEndSepPunct{\mcitedefaultmidpunct}
{\mcitedefaultendpunct}{\mcitedefaultseppunct}\relax
\EndOfBibitem
\bibitem[Hentschel \latin{et~al.}(2023)Hentschel, Koshelev, Sterl, Both, Karst, Shamsafar, Weiss, Kivshar, and Giessen]{mievoids}
Hentschel,~M.; Koshelev,~K.; Sterl,~F.; Both,~S.; Karst,~J.; Shamsafar,~L.; Weiss,~T.; Kivshar,~Y.; Giessen,~H. Dielectric Mie voids: confining light in air. \emph{Light, Science \& applications} \textbf{2023}, \emph{12}, 3\relax
\mciteBstWouldAddEndPuncttrue
\mciteSetBstMidEndSepPunct{\mcitedefaultmidpunct}
{\mcitedefaultendpunct}{\mcitedefaultseppunct}\relax
\EndOfBibitem
\bibitem[Galinski \latin{et~al.}(2017)Galinski, Fratalocchi, Döbeli, and Capasso]{hgalinski_lighmanipulation}
Galinski,~H.; Fratalocchi,~A.; Döbeli,~M.; Capasso,~F. Light Manipulation in Metallic Nanowire Networks with Functional Connectivity. \emph{Advanced Optical Materials} \textbf{2017}, \emph{5}, 1600580\relax
\mciteBstWouldAddEndPuncttrue
\mciteSetBstMidEndSepPunct{\mcitedefaultmidpunct}
{\mcitedefaultendpunct}{\mcitedefaultseppunct}\relax
\EndOfBibitem
\bibitem[Wohlwend \latin{et~al.}(2022)Wohlwend, Sologubenko, Döbeli, Galinski, and Spolenak]{wohlwend_chemicalengineering}
Wohlwend,~J.; Sologubenko,~A.~S.; Döbeli,~M.; Galinski,~H.; Spolenak,~R. Chemical Engineering of Cu–Sn Disordered Network Metamaterials. \emph{Nano Letters} \textbf{2022}, \emph{22}, 853--859, PMID: 34738817\relax
\mciteBstWouldAddEndPuncttrue
\mciteSetBstMidEndSepPunct{\mcitedefaultmidpunct}
{\mcitedefaultendpunct}{\mcitedefaultseppunct}\relax
\EndOfBibitem
\bibitem[Wohlwend \latin{et~al.}(2023)Wohlwend, Haberfehlner, and Galinski]{wohlwend2023}
Wohlwend,~J.; Haberfehlner,~G.; Galinski,~H. Strong Coupling in Two-Phase Metamaterials Fabricated by Sequential Self-Assembly. \emph{Advanced Optical Materials} \textbf{2023}, \emph{11}, 2300568\relax
\mciteBstWouldAddEndPuncttrue
\mciteSetBstMidEndSepPunct{\mcitedefaultmidpunct}
{\mcitedefaultendpunct}{\mcitedefaultseppunct}\relax
\EndOfBibitem
\bibitem[Karabacak and Lu(2005)Karabacak, and Lu]{enhancedstepcoverage}
Karabacak,~T.; Lu,~T.-M. Enhanced step coverage by oblique angle physical vapor deposition. \emph{Journal of Applied Physics} \textbf{2005}, \emph{97}, 124504--124504\relax
\mciteBstWouldAddEndPuncttrue
\mciteSetBstMidEndSepPunct{\mcitedefaultmidpunct}
{\mcitedefaultendpunct}{\mcitedefaultseppunct}\relax
\EndOfBibitem
\bibitem[Cremers \latin{et~al.}(2019)Cremers, Puurunen, and Dendooven]{conformalald}
Cremers,~V.; Puurunen,~R.; Dendooven,~J. Conformality in atomic layer deposition: Current status overview of analysis and modelling. \emph{Applied Physics Reviews} \textbf{2019}, \emph{6}, 021302\relax
\mciteBstWouldAddEndPuncttrue
\mciteSetBstMidEndSepPunct{\mcitedefaultmidpunct}
{\mcitedefaultendpunct}{\mcitedefaultseppunct}\relax
\EndOfBibitem
\bibitem[van~de Haar \latin{et~al.}(2016)van~de Haar, van~de Groep, Brenny, and Polman]{controlingmagneticandelectricdipolemodes}
van~de Haar,~M.~A.; van~de Groep,~J.; Brenny,~B.~J.; Polman,~A. Controlling magnetic and electric dipole modes in hollow silicon nanocylinders. \emph{Opt. Express} \textbf{2016}, \emph{24}, 2047--2064\relax
\mciteBstWouldAddEndPuncttrue
\mciteSetBstMidEndSepPunct{\mcitedefaultmidpunct}
{\mcitedefaultendpunct}{\mcitedefaultseppunct}\relax
\EndOfBibitem
\bibitem[Cao \latin{et~al.}(2009)Cao, White, Park, Schuller, Clemens, and Brongersma]{engineeringlightabsorpitioninsemicon}
Cao,~L.; White,~J.; Park,~J.-S.; Schuller,~J.; Clemens,~B.; Brongersma,~M. Engineering Light Absorption in Semiconductor Nanowire Devices. \emph{Nature materials} \textbf{2009}, \emph{8}, 643--7\relax
\mciteBstWouldAddEndPuncttrue
\mciteSetBstMidEndSepPunct{\mcitedefaultmidpunct}
{\mcitedefaultendpunct}{\mcitedefaultseppunct}\relax
\EndOfBibitem
\bibitem[Kühner \latin{et~al.}(2022)Kühner, Wendisch, Antonov, Bürger, Hüttenhofer, Menezes, Maier, Gorkunov, Kivshar, and Tittl]{outofplanesymmetrybreaking}
Kühner,~L.; Wendisch,~F.; Antonov,~A.; Bürger,~J.; Hüttenhofer,~L.; Menezes,~L.; Maier,~S.; Gorkunov,~M.; Kivshar,~Y.; Tittl,~A. Unlocking the out-of-plane dimension for photonic bound states in the continuum to achieve maximum optical chirality. \textbf{2022}, \relax
\mciteBstWouldAddEndPunctfalse
\mciteSetBstMidEndSepPunct{\mcitedefaultmidpunct}
{}{\mcitedefaultseppunct}\relax
\EndOfBibitem
\bibitem[Ghiradella(1991)]{structualcolors_butterflies}
Ghiradella,~H. Light and color on the wing: structural colors in butterflies and moths. \emph{Appl. Opt.} \textbf{1991}, \emph{30}, 3492--3500\relax
\mciteBstWouldAddEndPuncttrue
\mciteSetBstMidEndSepPunct{\mcitedefaultmidpunct}
{\mcitedefaultendpunct}{\mcitedefaultseppunct}\relax
\EndOfBibitem
\bibitem[Cao \latin{et~al.}(2022)Cao, Du, Guo, Hu, Zhang, Wang, Zhou, Gao, Fischer, Wang, Stavrakis, and deMello]{replicatingstructuralcolorsinbutterlies}
Cao,~X.; Du,~Y.; Guo,~Y.; Hu,~G.; Zhang,~M.; Wang,~L.; Zhou,~J.; Gao,~Q.; Fischer,~P.; Wang,~J.; Stavrakis,~S.; deMello,~A. Replicating the Cynandra opis Butterfly's Structural Color for Bioinspired Bigrating Color Filters. \emph{Advanced Materials} \textbf{2022}, \emph{34}, 2109161\relax
\mciteBstWouldAddEndPuncttrue
\mciteSetBstMidEndSepPunct{\mcitedefaultmidpunct}
{\mcitedefaultendpunct}{\mcitedefaultseppunct}\relax
\EndOfBibitem
\bibitem[Chen \latin{et~al.}(2018)Chen, Hou, Zhang, Pendry, and Chan]{Chen2018}
Chen,~W.-J.; Hou,~B.; Zhang,~Z.-Q.; Pendry,~J.~B.; Chan,~C.~T. Metamaterials with index ellipsoids at arbitrary k-points. \emph{Nature Communications} \textbf{2018}, \emph{9}, 2086\relax
\mciteBstWouldAddEndPuncttrue
\mciteSetBstMidEndSepPunct{\mcitedefaultmidpunct}
{\mcitedefaultendpunct}{\mcitedefaultseppunct}\relax
\EndOfBibitem
\end{mcitethebibliography}



\section*{Supplementary material}

# Supplementary Information: Hybrid resonant metasurfaces with configurable structural colors

Jelena Wohlwend,* Anna Hilti, Claudiadele Polinari, Ralph Spolenak, and Henning Galinski

*Laboratory for Nanometallurgy, Department of Materials, ETH Zurich, 8093 Zürich, Switzerland*

E-mail: jelena.wohlwend@mat.ethz.ch

## Large-scale angle independent structural colors with amorphous silicon aluminum brass

Due to their low density, aluminum brass commonly finds application in airspace. Here, however we explore their potential as backreflectors for large-scale angle independent structural colors by targeted chemical engineering of the refractive index, i. e. alloying. For this, we have deposited different aluminum brass compositions by physical vapor deposition (PVD) and determined their refractive indices through elipsometry. The influence of alloying on the refractive index of aluminum brass is discussed in the following section. The structural color is then generated by the additional deposition of an amorphous silicon (aSi) layer on top of the aluminum brass backreflector, schematically shown in Figure 1 a. The color response perceived by the human eye thereby stems from the combination of a non-trivial phase shift and phase accumulation through propagation in the lossy medium.[1] We thereby receive direct information on the silicon layer thickness through the color as the absorbing state of these architectures is determined by the film thickness.

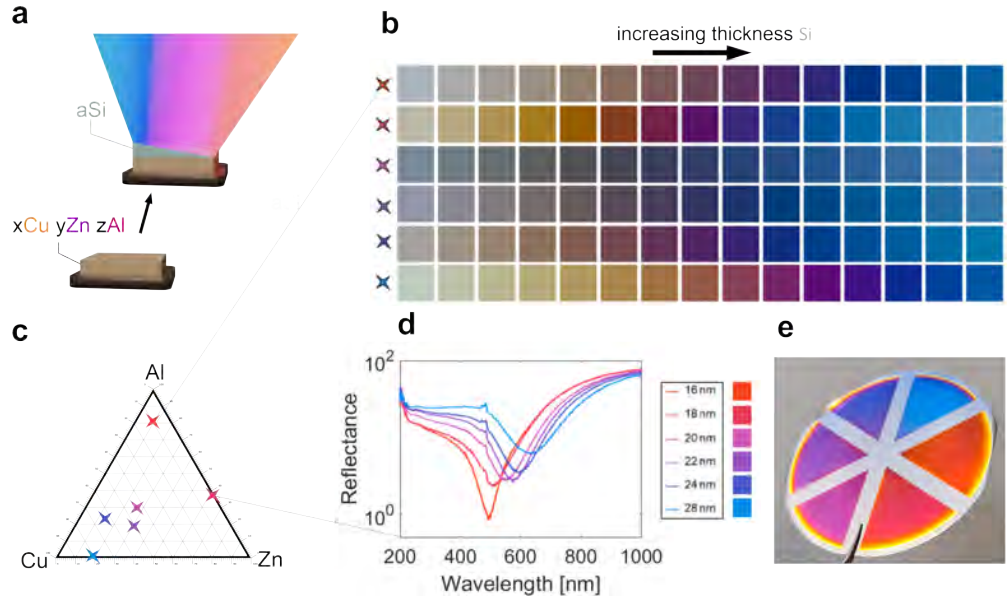

Figure 1: **Color engineering in ultra-thin optical coatings (a)** Schematic illustration of the two layer optical coating with an aluminum brass backreflector of different compositions and an amourphos silicon layer in different thickness ontop of it **(b)** Simulated color arising from the deposition of amorphous silicon (in increments of 2 nm going to the right) on different aluminum brass alloys, indicated with colored stars on the right. **(c)** Phase diagram of the selected aluminum brass backreflectors. The colored stars represent the selected compositions. **(d)** Near-normal incident reflectance of a selected sample with a red shift of the absorbing state with increasing thickness and **(e)** photograph of the selected sample highlighting the color change with increasing Si thicknesses.

The influence of chemically engineering the backreflector is shown in Figure 1 b. There the simulated color arising from a thin silicon layer on different backreflectors are shown. The calculated colors are displayed in ascending thicknesses of 2 nanometers from left to right, with the first colored square representing the color of the substrate itself (0 nanometers of aSi). The colored symbol on the left side represents the aluminum brass composition (Figure 1 c). Different colors are observed ranging from brown to yellow over pink and violett to blue depending on the backreflector as well es the thickness of the a-Si layer. In later experiments with the a-Si nanocups only the 82 % Al, 9 % Cu and 9 % Zn was used, to fulfill compositional requirements for the chemical dealloying process.[2] Figure 1 d shows near normal incident reflectance spectra of different Si thicknesses on a selected backreflector. With increasing thickness the absorbing state shifts towards the near IR resulting in a blue

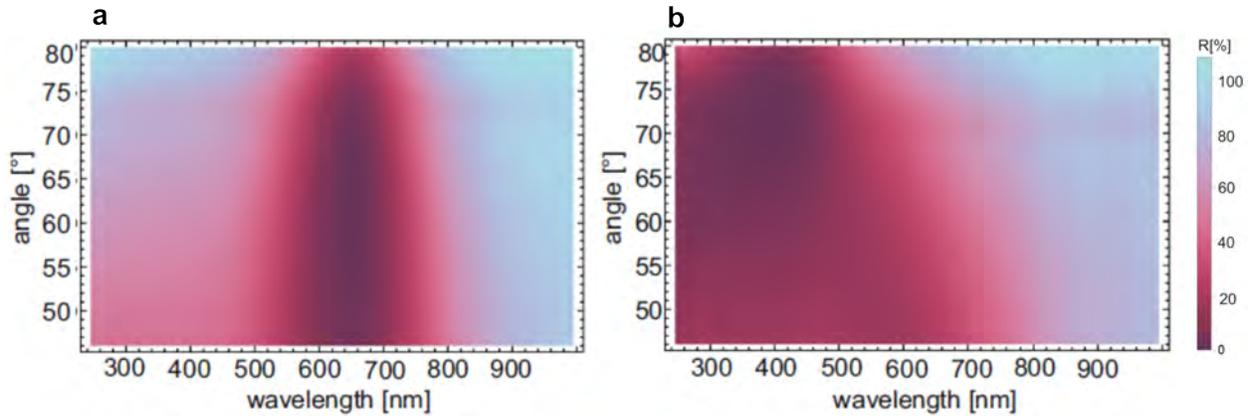

Figure 2: **Angle independent colors:** Angle-dependence reflectance of a selected 28 nanometer silicon film on an brass backreflector for **(a)** s-polarized light and **(b)** p-polarized light

coloration. An optical image of the sample is shown in Figure 1 e. A variety of colors starting from orange over pink to blue are generated depending on the thickness of the a-Si layer. Angle dependent reflectance measurements are presented in Figure 2. The dark parts of the colormap plots show low reflectance, denoting near-perfect absorbance of a specific wavelength of light with a certain incident angle. For s-polarized light (Figure 2 a) for wavelengths of 590-799 nanometers the vertical bands of absorption are clearly visible. As they stretch fully throughout the graph (reflectance minimum is persistent for angles of incidence between 45° and 80°), the absorption and thus generated color is independent of the angle of incident. The graph for p-polarized light (Figure 2 b) does not display the same behaviour as for s-polarized light. P-polarized light should in fact not be regarded when considering angle-independency, as it is subject to Brewster's angle. At this angle, which varies for every material, there is no reflection of p-polarized light. Therefore, the detected reflection minimum at a given incident angle could be the result of near-perfect absorption by the silicon film or no reflection (if the incident angle corresponds to the Brewster's angle).

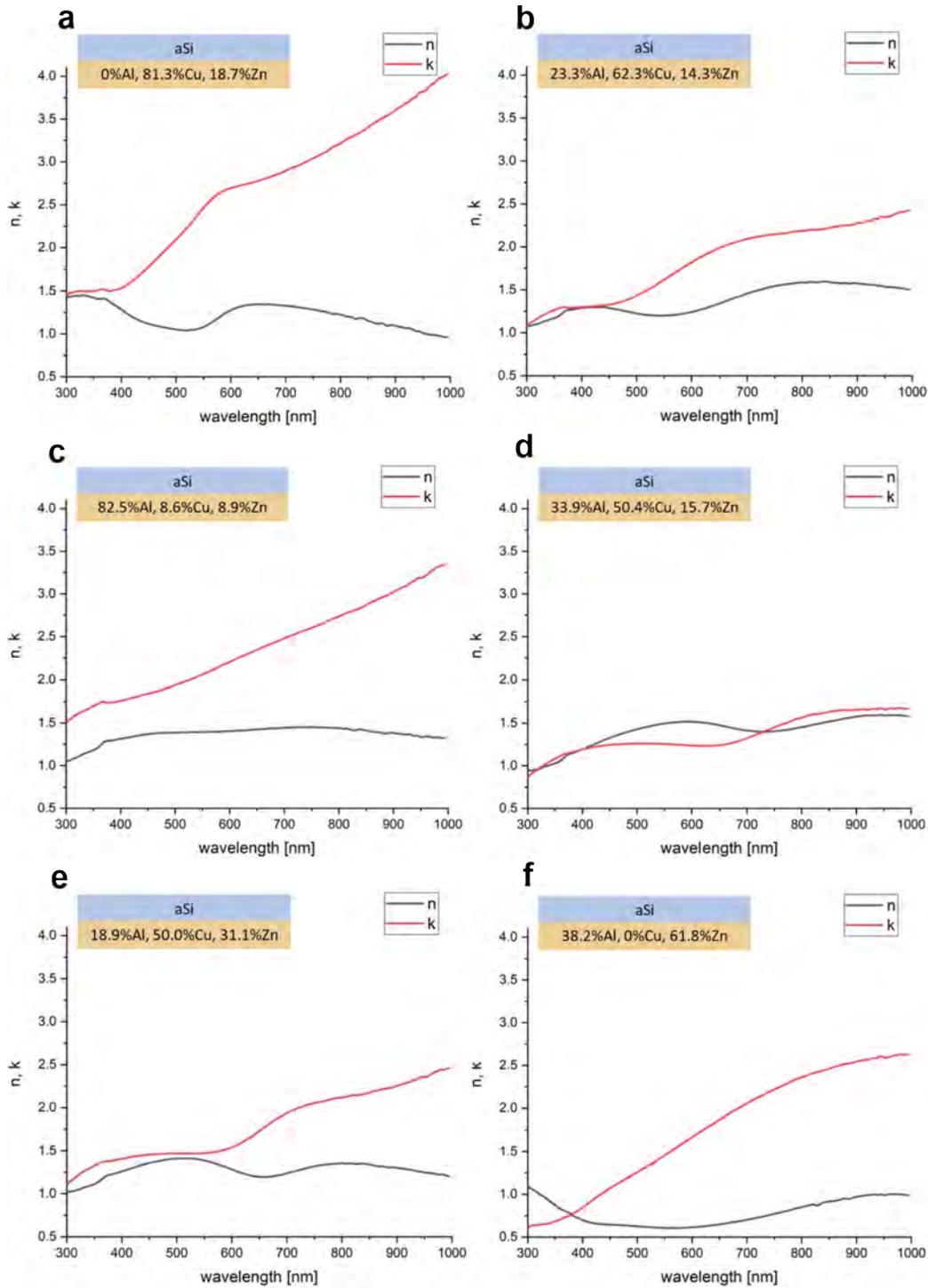

Figure 3: **Refractive index of the Al-Cu-Zn system** Measured refractive index of different Al-Cu-Zn alloys for different compositions denoted in the schematics

## Refractive indices of different aluminum brass

The frequency-dependent complex refractive index $n$ and $k$ from the backreflector were measured and are presented in Figure 3. By changing the composition of the alloy, the

complex refractive index, can be engineered.[3,4] Our measurements confirm this and show that, except the backreflector in 3 d, the aluminum brass behaves metallic ($n < k$) in the visible regime indicating that the all light is reflected. No clear correlation between composition and the change in the refractive index can be observed.

## Directed growth of the a-Si nanocup metasurface

Figure 4 illustrates the sputtering mechanisms and details leading to the formation of amorphous silicon (a-Si) nanocups. Material deposition for physical vapor deposition is limited to line-of-sight deposition (shown in Figure 4 a). A tilt ($< 90°$) of the sputtering source can therefore result in wall coverage of trench like structures. Furthermore, in our case, the deposition kinetics result in non-uniform growth conditions with a thickness gradient of the sidewall width $s_{sw}$ (see Figure 4 b, where an incomplete lift-off is shown). This thickness gradient results in a predetermined breaking point (Figure 4 a, yellow star) during the lift off. In Figure 4 b an incomplete lift-off is shown illustrating the predetermined breaking point as well as the non-uniform growth.

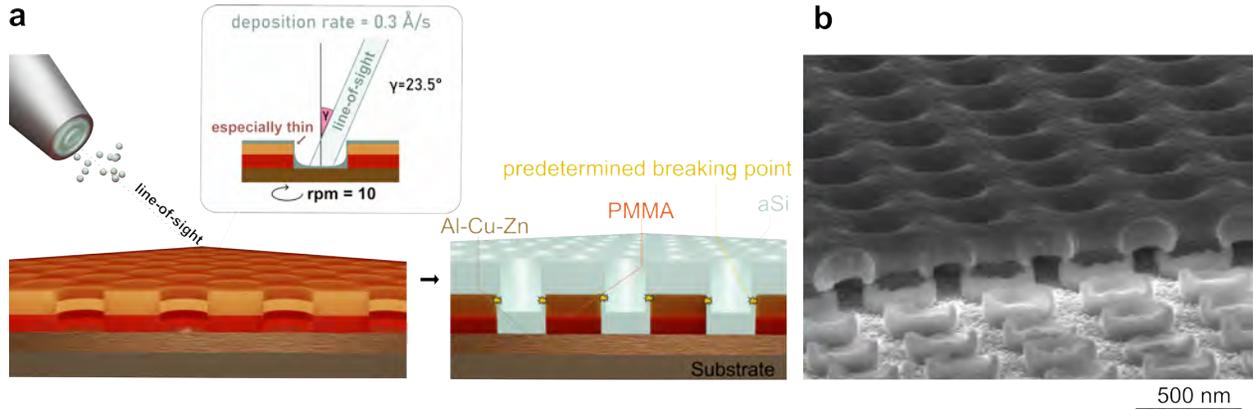

Figure 4: **Directed growth of a-Si nanocup metasurface (a)** Schematic illustration of the sputtering process showing the direct growth of the a-Si nanocups with the inlet showing fabrication details **(b)** incomplete lift-of of deposited Si layer, outlining the formation mechanism.

## Optical response of a-Si nanocups on $Si/SiO_2$

Figure 5 a depicts an optical micrograph of a-Si cups with different diameters and periodicities grown on $Si/SiO_2$. Reflectance curves of the different silicon nanocups are shown in 5 b. Similar to the nanocups on aluminum brass a blue shift of the reflectance minima is observed with increasing cup diameter. Additionally, also a broadening of the minima with increasing cup diameter is observed (Figure 5 b). In Figure 5 c scanning electron micrographs of a-Si nanocups on Si/SiO₂ with different diameters and periodicities are presented.

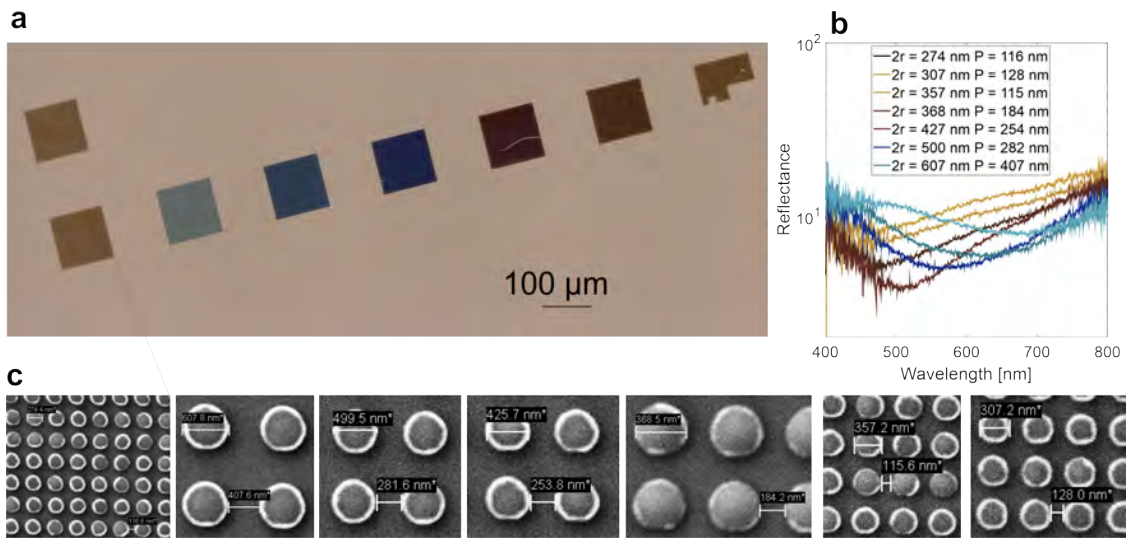

Figure 5: **Optical response of a-Si nanocups on $Si/SiO_2$ (a)** optical micrograph of a-Si cup metasurfaces with different diameters **(b)** Reflectance measurements of a-Si cups with different diameters and **(c)** scanning electron micrographs of a-Si nanocups on Si/SiO₂ with different diameters.

## Color engineering by tailoring the geometrical dimensions of the a-Si nanocups

In Figure 6 the influence of different geometrical parameters are shown. The different parameters were investigated through finite element simulations of a 2D projection of the a-Si nanocups, namely cup diameter $2r$, base height $h_b$, width side wall $w_{sw}$ and the periodicity $p$, i. e. spacing between the cups (Figure 6 a). Details on the simulations are given in the

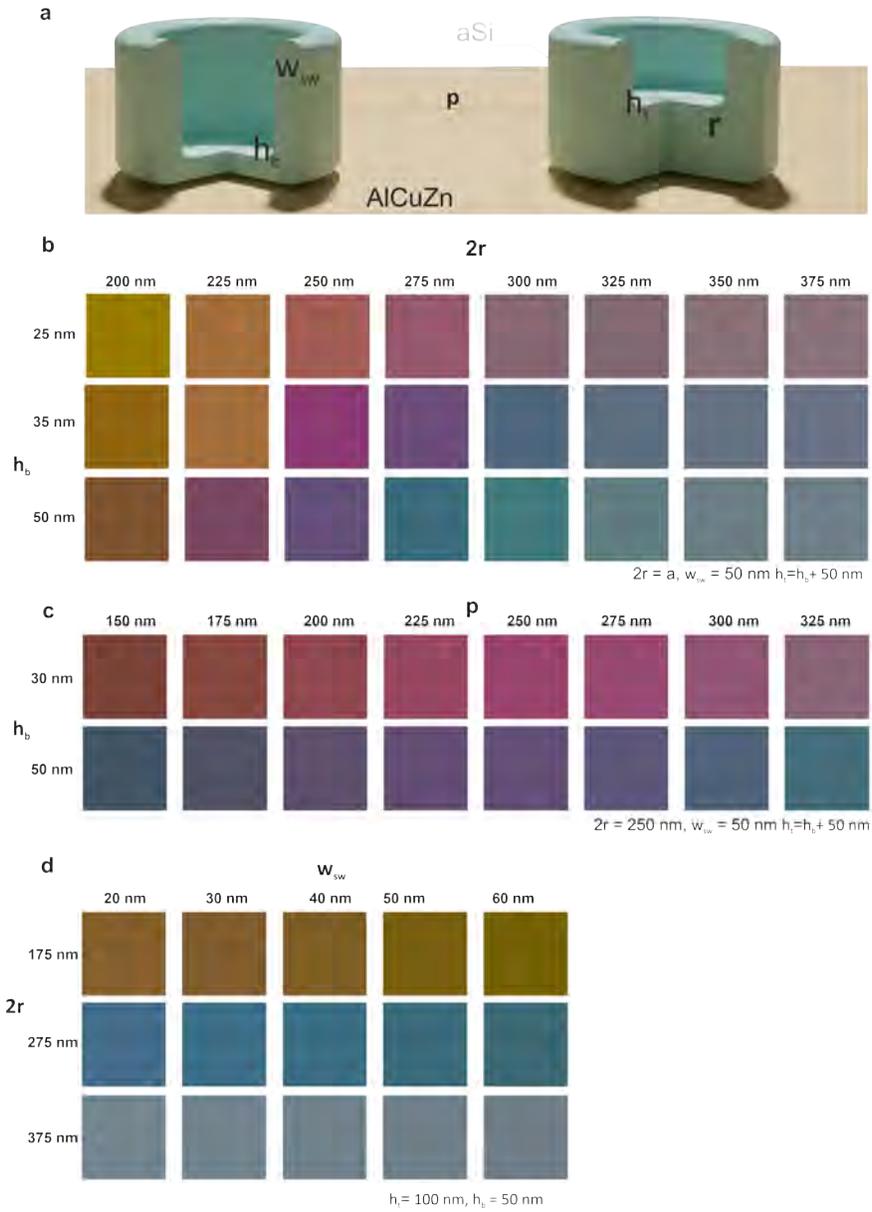

Figure 6: **Simulated color engineering:** (a) Schematic illustrating the investigated geometrical parameters and (b) simulated influence of base height $h_b$ and cup diameter (2r) on the perceived color. (c) Influence of the periodicity and (d) influence of the side wall width ($w_{ws}$) on the resulting color

experimental section.

Firstly the influence of an increasing cup diameter is shown in Figure 6 b. With increasing cup diameter (2r) the color transitions from yellow to orange to pink and purple and finally blue. Also the base height strongly influences the perceived color, shown in Figure 6 c. An

7

increasing base height shifts the color to blue for smaller diameters. Through the comparison of the experimentally observed and simulated colors it is indicated, that an increase in cup diameter also results in an increase in base height. This means the simulation match the experiments best if it is assumed that an increase in diameter is accompanied by an increase in base height. Intuitively, this can be seen as more material is deposited in a trench for larger trench diameters. An increase in periodicity results in slight changes of the chroma and hues of the simulated colorss (Figure 6). The increase in periodicity mainly influences the lattice mode of the nanocups discussed in the next section. Lastly, the influence of the side wall width is shown in Figure 6 d. The sidewall width shows only a very small influence on the resulting colors of the a-Si nanocup metasurfaces.

## Lattice Mode in Nanocup Metasurfaces

In Figure 7 a the simulated reflectance for different periodicities p are shown. With increasing periodicities, the lattice mode shifts towards the red (Figure 7 b). Furthermore, in Figure 7 c the normalized electric field distribution is showing the formation of lattice modes confined between different cups with increasing lattice constants p.

## Scattering behaviour of Si nanocups

In Figure 8 a the comparison of the forward and backward scattering for an a-Si nanocup with diameter 2r= 460 nm, base height $h_b$=30nm and wallthickness $w_s w$=50nm is shown additionally to the scattering crosssection for different wavelength in Figure 8 b. Predominately forward scattering is observed, this is especially interesting for the interaction of a-Si nanocups with the substrate.

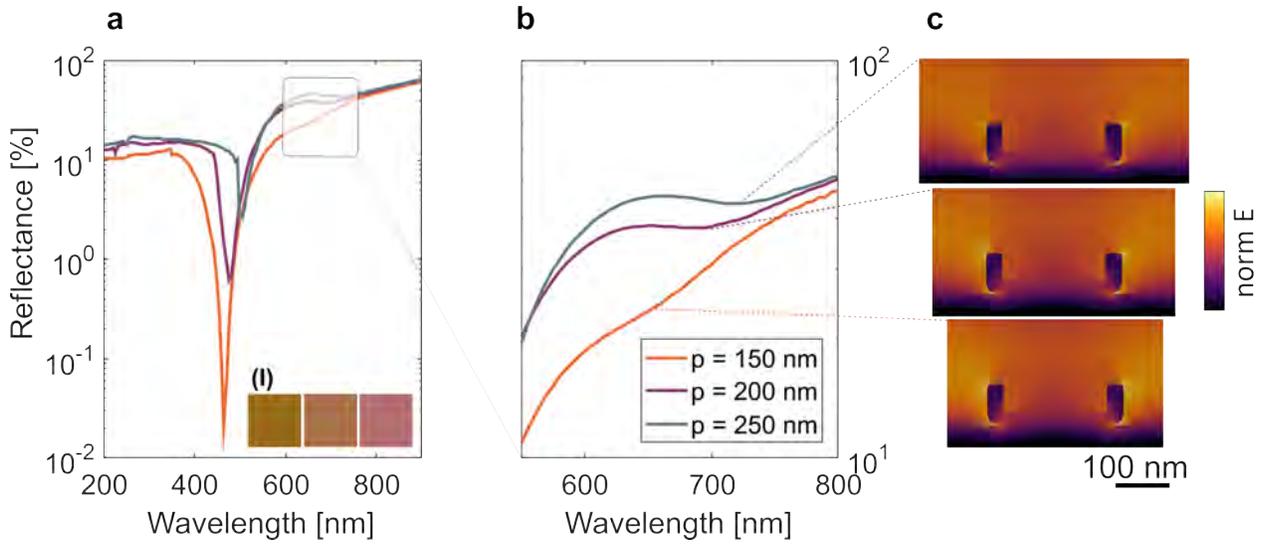

Figure 7: **Lattice modes of a-Si nanocup metasurfaces (a)** Simulated reflectance with increasing periodicity p with (I) showing the simulated colors. **(b)** Enlarged cut out showing only the lattice mode for different periodicities and **(c)** normalized electric field showing the formation of a lattice mode concentrated between two cups.

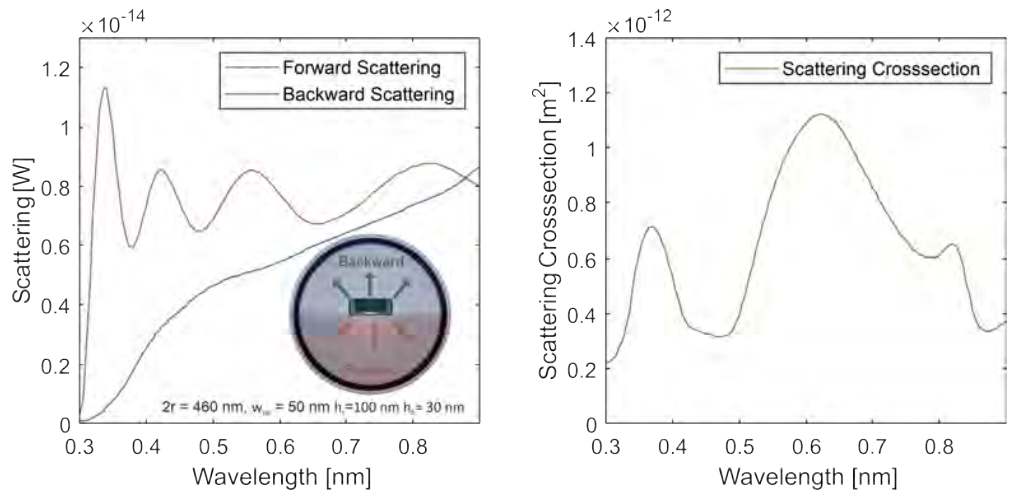

Figure 8: **Scattering behaviour of Silicon Nanocups (a)** Simulated wavelength dependent forward and backward scattering and **(b)** simulated scattering cross section for different wavelength.

## Non-reciprocity of Si nanocups

In Figure 9 the non-reciprocity, i. e., the simulated directional dependent response of the a-Si nanocup metasurface is visualized. Where as the transmission is only slightly different, the reflectance for the two different excitation directions (from the top and from the buttom

9

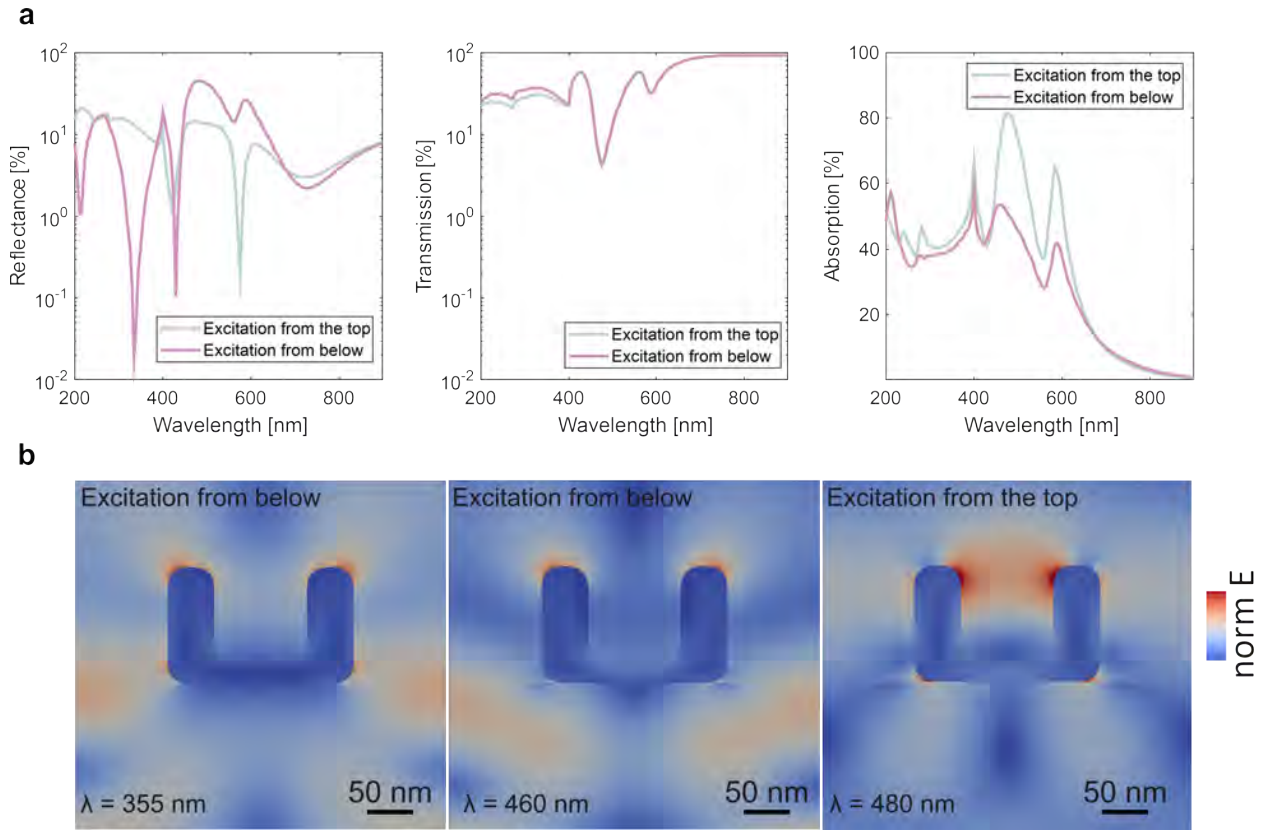

Figure 9: **Non-reciprocity (a)** From left to right simulated directional dependent reflectance, transmission and absorption of a-Si nanocup on $SiO_2$ and **(b)** normalized electric field distribution showing the formation different modes depending on the excitation directions.

of the 2D simulation) significantly deviates from each other. Through a change in excitation direction, also , charactericed by a reflectance minima. The normalized electric field distribution of selected modes is shown in Figure 9 b.

## Configurable structural colors through order disorder transition

In Figure 10 the interaction between different nanocup arrays and the disordered network metamaterial (DNM) are shown. Chemical deallyoing, i. e. the introduction of disorder through the formation of a disordered network instead of the backreflecter, meaning also the transition from a reflective to an absorptive substrate shifts the reflectance minima towards the red for all combinations of nanocups and DNMs, visible by comparing Figure 10 a and

10

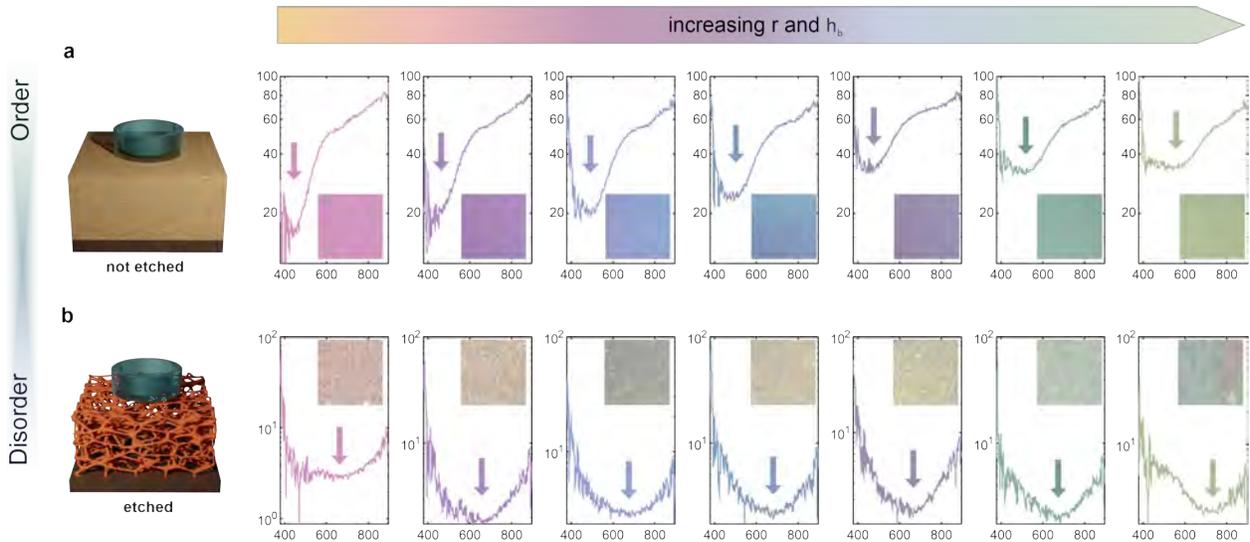

Figure 10: **Configurable structural colors through order disorder transition** The schematic on the left depicts the state of the system, ordered for **(a)** and disordered for **(b)**. Furthermore, the reflectance spectra (with increasing r and $h_b$) for both the ordered and the disordered state are shown on the right. The colored subfigures in the reflectance plots depict optical micrographs of the different nanocup metasurface underlining the color change between the two states.

b. This shift in the resonant modes indicates mode coupling between the nanocup arrays and the DNMs.

# References

(1) Kats, M.; Capasso, F. Optical absorbers based on strong interference in ultra-thin films. *Laser & Photonics Reviews* **2016**, *10*.

(2) Wohlwend, J.; Sologubenko, A. S.; Döbeli, M.; Galinski, H.; Spolenak, R. Chemical Engineering of Cu–Sn Disordered Network Metamaterials. *Nano Letters* **2022**, *22*, 853–859, PMID: 34738817.

(3) Littleton, J. T. The Optical Constants of Alloys as a Function of Composition. *Physical Review (Series I)* **1911**, *33*, 453–466.

(4) Das, U.; Bhattacharya, P. K. Variation of refractive index in strained InxGa1-xAs-GaAs heterostructures. *Journal of Applied Physics* **1985**, *58*, 341–344.



\end{document}